\documentclass[conference,twocolumn]{IEEEtran}
\IEEEoverridecommandlockouts

\usepackage{longtable}
\usepackage{cite}
\usepackage{amsmath,amssymb,amsfonts}
\usepackage{algorithmic}
\usepackage{graphicx}
\usepackage{textcomp}
\usepackage{hyperref}
\usepackage{xcolor}
\def\BibTeX{{\rm B\kern-.05em{\sc i\kern-.025em b}\kern-.08em
    T\kern-.1667em\lower.7ex\hbox{E}\kern-.125emX}}
\begin{document}

\title{ Evaluation Metrics for Misinformation Warning Interventions: Challenges and Prospects
\vspace{10px}
[file:////255b1em]\\[1em] \normalsize{\textbf{Author's draft for soliciting feedback - \today}}

}

\title{ Evaluation Metrics for Misinformation Warning Interventions: Challenges and Prospects\\

\vspace{10px}

\normalsize{\textbf{\textcolor{red}{Author's draft for soliciting feedback - \today}}}

\author{
 
    Hussaini Zubairu, Abdelrahaman Abdou\textsuperscript{1}, and	Ashraf Matrawy \\ School of Information Technology, Carleton University, Canada \\ \textsuperscript{1}School of Computer Science, Carleton University, Canada 
   
	}

}

\maketitle

\begin{abstract}
Misinformation has become a widespread issue in the 21\textsuperscript{st} century, impacting numerous areas of society and underscoring the need for effective intervention strategies. Among these strategies, user-centered interventions, such as warning systems, have shown promise in reducing the spread of misinformation. Many studies have used various metrics to evaluate the effectiveness of these warning interventions. However, no systematic review has thoroughly examined these metrics in all studies. This paper provides a comprehensive review of existing metrics for assessing the effectiveness of misinformation warnings, categorizing them into four main groups: behavioral impact, trust and credulity, usability, and cognitive and psychological effects. Through this review, we identify critical challenges in measuring the effectiveness of misinformation warnings, including inconsistent use of cognitive and attitudinal metrics, the lack of standardized metrics for affective and emotional impact, variations in user trust, and the need for more inclusive warning designs. We present an overview of these metrics and propose areas for future research.
\end{abstract}


\section{Introduction}

The reliance on social media as a primary source of news, coupled with the absence of editorial oversight, has elevated misinformation to a global concern~\cite{kirchner2020countering}. Misinformation refers to false, incorrect, inaccurate, misleading, or out-of-context information, regardless of the intent behind its dissemination~\cite{wu2019misinformation}~\cite{rastogi2023review}. Although misinformation in various forms has existed for a long time~\cite{allcott2017social}, its propagation has recently intensified with the advent of modern digital technologies~\cite{sharma2019combating}.

The impacts of misinformation affect all aspects of human existence, including political, social ~\cite{sharma2019combating}\cite{shu2019defend}, and economic~\cite{singhal2023cybersecurity}. Given these profound effects, misinformation has become an important concern due to its influence on public opinion~\cite{jahanbakhsh2021exploring}, as evidenced in political landscapes~\cite{shu2019defend} and behavioral attitudes during the COVID-19 pandemic~\cite{ling2023learn}.

The widespread prevalence of misinformation has prompted social media platforms to implement warning interventions aimed at mitigating its impact~\cite{jahanbakhsh2021exploring}. The primary goal of these warnings is to serve as critical interventions to curb the spread of misinformation while encouraging users to critically evaluate content before sharing~\cite{jahanbakhsh2021exploring}. Misinformation warnings are cautionary notifications that may take various forms, including labels~\cite{kirchner2020countering, kaiser2021adapting}, symbols~\cite{sharevski2022meaningful}, pop-ups~\cite{greenberg2024heed}, or browser extensions~\cite{jahanbakhsh2024browser} that are attached to digital content of questionable credibility.

These warnings aim to inform users and influence their decisions about sharing such information. However, the effectiveness of misinformation warnings remains a contested topic in the literature. While some studies report positive outcomes~\cite{kaiser2021adapting,jahanbakhsh2021exploring}, others suggest limited impact on user behavior~\cite{grady2021nevertheless,capewell2024misinformation,van2023can}. A significant factor contributing to this inconsistency is the lack of standardized metrics for evaluating the effectiveness of these interventions.
Existing literature examines various aspects of misinformation interventions, including their effectiveness, design, and broader societal impacts. Tay et al.,~\cite{tay2023focus} propose comprehensive frameworks for evaluating misinformation interventions, emphasizing the need for an evaluation framework that assesses intervention effectiveness and its wider implications, such as polarization, skepticism, and fostering norms that value truth. Hartwig et al.,~\cite{hartwig2024landscape} propose taxonomies for evaluating interventions but focus on long-term impacts, user engagement, and transparency rather than specific performance metrics.

Similarly, Janmohamed et al.,~\cite{janmohamed2021interventions}, in their review, highlight the effectiveness of fact-checking and warning interventions; however, their emphasis lies on short-term outcomes and cultural applicability rather than on evaluation metrics. Smith et al.,~\cite{smith2023systematic} review various misinformation intervention strategies, their effectiveness, and the challenges associated with their assessment. While their study identifies widely used measures such as perceived accuracy, willingness to share misinformation, sharing discernment, and intent to vaccinate, it emphasizes the importance of standardized outcome measures without specifically focusing on reviewing metrics.

These works show that apart from the metrics mentioned by Smith et al.,~\cite{smith2023systematic}, no other study has comprehensively examined metrics for misinformation interventions in detail. To our knowledge, no previous review has explicitly focused on reviewing metrics for misinformation warning interventions. Our research uniquely addresses this gap and suggests the need to develop a unified and standardized framework to evaluate the performance of warning interventions. The objective is to establish a consistent basis for assessing the effectiveness of these interventions across platforms and contexts.

This paper addresses this gap by reviewing existing evaluation metrics in the literature. We aim to identify key challenges in current evaluation practices and propose opportunities for future research in this domain. We aim to address the following research questions (RQs):

\vspace{10px}
RQ1. What are the existing metrics for assessing the effectiveness of misinformation warnings? 

\vspace{10px}

\vspace{10px}
RQ2. What are the challenges and opportunities of existing metrics to measure the effectiveness of misinformation warnings?

\vspace{10px}

We chose to pursue these research questions due to the pressing need to identify reliable metrics and establish measurable and impactful standards for evaluating misinformation warning interventions. This review aims to uncover gaps in existing warnings interventions and explore innovative strategies to enhance the effectiveness of such interventions. Collectively, these questions are intended to contribute to developing more impactful and evidence-based warning intervention strategies.

The contributions of this paper are as follows.

\begin{enumerate}
  \item We reviewed the metrics used to assess the effectiveness of misinformation warnings, providing a detailed understanding of their assessment methods, and highlighting their inherent limitations (Section III).
  
  \item We developed a classification framework for existing metrics, organizing them into four broad categories: behavioral, trust and credibility, usability, and cognitive and psychological factors (Section III).
  
  \item Based on the gaps identified in the literature, including the lack of a standardized metrics framework, the challenges with accessibility and usability metrics, and the need for inclusive misinformation warning interventions, we suggest that research focus on developing more reliable, consistent and measurable metrics to evaluate the effectiveness of misinformation warning interventions (Section IV).
\end{enumerate}
\section{Research Methodology}

We adopted the Preferred Reporting Items for Systematic Reviews and Meta-Analyses (PRISMA) framework to guide our review process~\cite{hartwig2024landscape}. Relevant articles were identified through comprehensive searches in ScienceDirect, ACM Digital Library, IEEE Xplore, Google Scholar, and conference proceedings from the ACM Conference on Human-Computer Interaction. The search used the keywords 'misinformation', 'warning', 'intervention', 'metrics', and 'effectiveness', along with their Boolean combinations.

No restrictions were placed on the publication year, and studies up to 2024 were included. In addition to database searches, we employed backward and forward citation tracking to identify relevant studies that may have been missed in the initial search. This review's inclusion and exclusion criteria are presented in Table \ref{tab:tab1}.

\begin{table*}
\caption {Study Inclusion and Exclusion Criteria }
\label{tab:tab1}
\centering
\begin{tabular}{|p{0.5cm}|p{6cm}|p{6cm}|}
\hline
\textbf{S/N} & \textbf{Inclusion }& \textbf{Exclusion}  \\ \hline
1  & Peer-reviewed journal articles, conference proceedings & Non-peer-reviewed articles (blog posts, opinion pieces) \\ \hline
2  & Studies focusing on the evaluation of misinformation warnings intervention on social media platforms & Studies focusing solely on misinformation detection without addressing intervention or warnings  \\ \hline

3  & Articles that used metrics related to behavioral, cognitive, psychological, or usability aspects of warning intervention & Article is not published in a reputable research society, and not peer-reviewed. \\ \hline
4  & Studies published in English up to 2024 & Articles focus on misinformation interventions but lack discussion of performance evaluation metrics \\ \hline

\end{tabular}
\end{table*}

\section{Existing Metrics for Assessing Warnings Intervention}

\begin{figure}[htbp]
\centerline{\includegraphics[width=8cm,height=6cm]{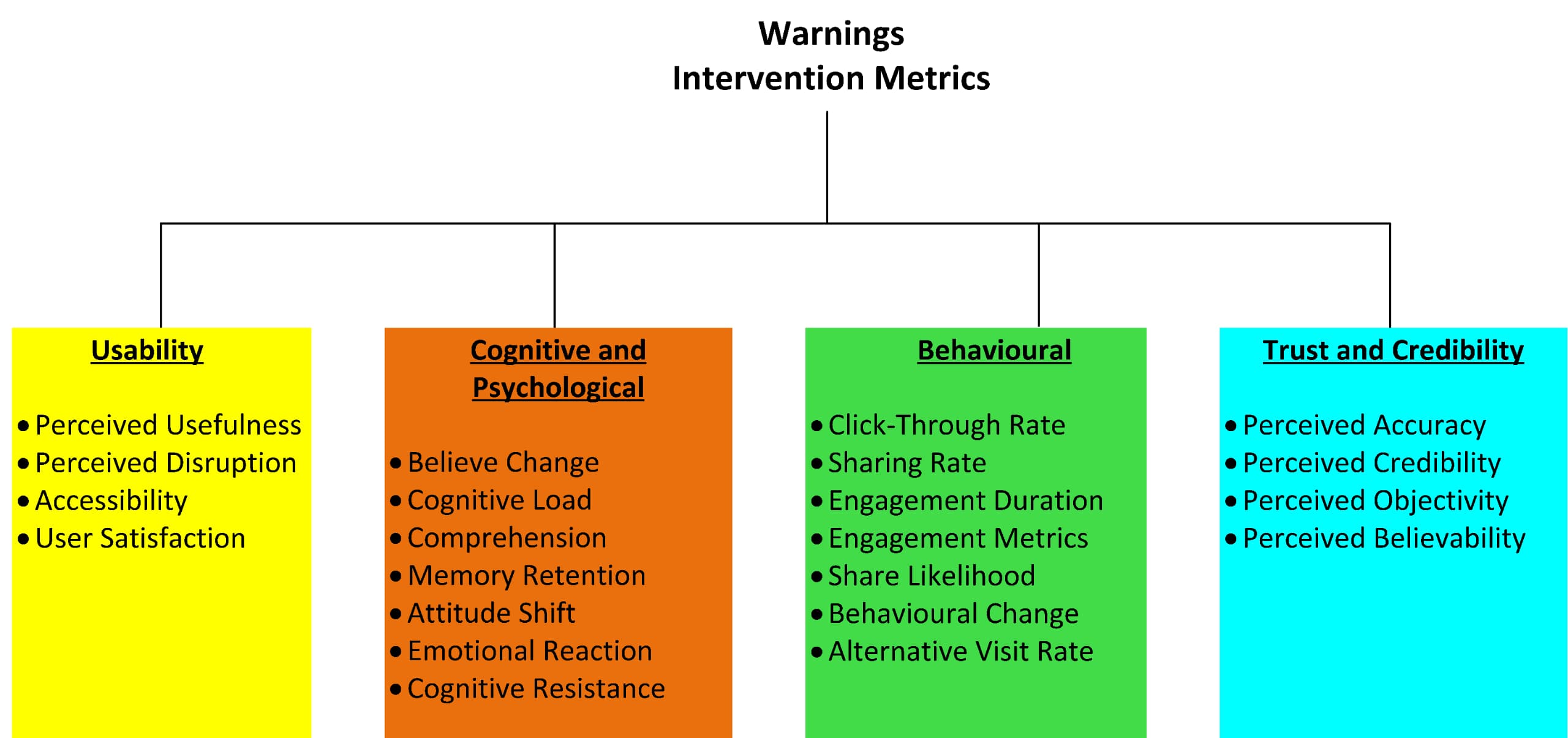}}

\caption{Proposed Classification of  Metrics
}
\label{fig1}
\end{figure}

As misinformation continues to pose a significant challenge, exacerbated by the proliferation of social media, widespread smartphone availability, and affordable internet access, warning interventions have emerged as a critical strategy to mitigate the spread of misinformation across social media platforms and online sources. To assess the effectiveness of these interventions, the existing body of literature employs a variety of evaluative metrics, which we have broadly categorized into four general domains: behavioral, trust and credulity, usability, and cognitive and psychological metrics, as illustrated in Figure \ref{fig1}.
 
\subsection{Behavioral Metrics}

As employed in existing studies, behavioral metrics focus on users' responses to warning interventions, offering researchers valuable insights into the impact of these warnings on user behavior. These metrics are essential for evaluating the effectiveness of warning interventions in influencing user behavior. In the present classification, this category encompasses the following metrics:

\begin{itemize}
    \item Click-Through Rates (CTR): This metric is employed to assess how frequently users interact with a warning intervention by clicking on it~\cite{kaiser2021adapting}\cite{guo2023seeing}.  Conceptually, a high CTR may indicate the effectiveness of the warning in capturing users' attention~\cite{kaiser2021adapting}\cite{guo2023seeing}. However, while this metric reflects user engagement and interest, it does not necessarily signify a change in users' attitudes or behaviors following exposure to the warning intervention. 

  \item Alternative Visit Rate (AVR): The AVR metric assesses the frequency with which social media users seek alternative information sources after encountering a misinformation warning~\cite{kaiser2021adapting}. It provides insight into whether the intervention prompts users to critically assess the veracity of the information and seek other perspectives as a result of the warning. This metric offers valuable information on how effectively the intervention encourages users to critically evaluate content and consider different viewpoints, which is crucial in mitigating the spread of misinformation. Kaiser et al.,~\cite{kaiser2021adapting} employed this metric to measure the impact of the warning intervention relative to the baseline rate. 
  
  \item Sharing rates: This metric is widely employed in the existing literature to assess the behavioral impact of misinformation warnings by analyzing the patterns of content sharing following a user's encounter with such warnings. Sharing rates are straightforward to quantify and provide direct insights into how warnings influence user actions. For instance, studies by Jahanbakhsh et al.,~\cite{jahanbakhsh2021exploring} and Jia et al.,~\cite{jia2022understanding} have utilized this metric to evaluate the effectiveness of warning strategies. While this measure offers valuable insights into short-term behavioral changes, its primary focus on a specific action (i.e., sharing) neglects other user behaviors and factors that may also influence the decision to share content, representing a notable limitation.
  \item Sharing likelihood: Metrics for sharing likelihood evaluate the probability that a user will share misinformation after encountering a warning~\cite{jahanbakhsh2021exploring}. This metric directly impacts the potential virality of misinformation across various social media platforms. Studies such as Jahanbakhsh et al.,~\cite{jahanbakhsh2021exploring} and~\cite{lees2022twitter} employed this metric to assess the effectiveness of warnings on users' sharing behavior through both behavioral data and self-reported measures. Reducing the likelihood of sharing is considered a key indicator of the intervention's effectiveness. 
  
  These studies measure this metric through participant responses, where participants were asked whether they would share a given piece of content on social media, selecting from the options: “Yes,” “Maybe,” or “No.” Researchers then analyzed these responses to identify factors influencing the likelihood of sharing.
  
  \item Engagement duration: Engagement duration metrics are used to assess the amount of time users spend interacting with warning interventions. This metric is based on the assumption that more extended engagement implies the effectiveness of the warning in capturing users' attention, potentially leading to critical thinking and behavioral change. For example, Greenberg~\cite{greenberg2024heed} employed this metric to evaluate the impact of warnings by measuring engagement duration. However, the underlying assumption of this metric may not always hold; for instance, a longer engagement duration could result from user confusion or skepticism. This metric does not account for engagement quality, such as whether users critically assess the information or merely skim the content. 
  \item Engagement metrics: Engagement metrics assess users' interaction with misinformation warnings, typically encompassing likes, comments, retweets, and shares~\cite{guo2023seeing}\cite{greenberg2024heed}. Researchers monitor these metrics to evaluate the effectiveness of the intervention; a high level of engagement often indicates that the intervention is successfully encouraging user interaction. Existing studies, such as those by Guo \textit{et al.,}~\cite{guo2023seeing}, Green~\cite{greenberg2024heed}, and Sharevski \textit{et al.,}~\cite{sharevski2022meaningful}, have employed various strategies to measure these metrics to gain insights into the impact of warning interventions on mitigating misinformation. 
   \item Behavioral change: The behavioral change metric assesses alterations in individuals' decision-making processes following exposure to misinformation warnings~\cite{seo2024reliability}. In the context of misinformation, it serves as an essential tool for evaluating the effectiveness of warning interventions and is considered strong evidence of an intervention's success~\cite{seo2024reliability}. Behavioral change can be measured through observation, recording, or analysis of user responses, with the method selected depending on the nature of the behavior, the environment, and the desired outcomes. For instance, Seo et al.,~\cite{seo2024reliability} employed this metric to evaluate users' sharing behavior after exposure to misinformation warnings on social media, finding a reduction in the dissemination of misinformation, indicating a positive behavioral change~\cite{seo2024reliability}.
\end{itemize}

The advantages and limitations of commonly used behavioral metrics in the literature are summarized in Table \ref{tab:tab2}.

\begin{table}
\caption {Commonly Employed Behavioral Metrics}
\label{tab:tab2}
\centering
\begin{tabular}{|p{1.5cm}|p{2.4cm}|p{2.5cm}|}
\hline
\textbf{Metrics} & \textbf{Pros }& \textbf{Cons} \\ \hline
Click-Through Rates  & Indicates user engagement & Platform specific \\ \hline
Sharing behavior  & Directly reflects behavior & Sharing may occur for other reasons  \\ \hline

Engagement (likes, comments)  & Useful for sentiment analysis & Ambiguity \\ \hline
Alternative Visit Rate (AVR)  & Reveal openness to alternative source & Platform specific \\ \hline

Sharing rates & Easily to measure & It can be influenced by other factors\\ \hline

Sharing likelihood  & Useful to measure individual predisposition to share after exposure  & It may not reflect real-world behavior\\ \hline 
\end{tabular}
\end{table}

Generally, while these metrics capture observable user actions and offer direct evidence of an intervention’s impact, behavioral metrics may be insufficient to fully reflect the complexity of user responses to warning interventions. Table II shows that metrics such as sharing rates, likelihood of sharing, and engagement indicators may suggest the effectiveness of warnings. Still, they do not necessarily indicate changes in users’ underlying beliefs. Similarly, engagement metrics can be ambiguous, as engagement and lack of engagement may result from factors unrelated to the warning intervention~\cite{clayton2020real}. The context-specific nature of specific behavioral metrics underscores the need for complementary measures that capture deeper cognitive and attitudinal processes alongside observable user behavior.

\subsection{Trust and Credibility Metrics}

Trust plays a crucial role in how people engage with information. Therefore, evaluating the effectiveness of an intervention involves measuring how it impacts the user's trust in either the information or the warning itself. Trust and credibility metrics, therefore, assess the impact of intervention on users’ perceptions as it relates to the credibility, reliability, and integrity of both the warning and the source of the warnings. In the present classification, this category encompasses the following metrics:
\begin{itemize}
    \item Perceived accuracy: Accuracy perception is one of the most commonly used metrics to evaluate the effectiveness of misinformation warnings. It assesses how users perceive the reliability of information after being exposed to a warning. Researchers, including Kirchner and  Reuter~\cite{kirchner2020countering},  Jahanbakhsh et al.,~\cite{jahanbakhsh2021exploring}, and Donabauer et al.,~\cite{donabauer2024empowering}, among others, have investigated the impact of perceived accuracy on user behavior, revealing varying degrees of accuracy perception among users. This metric is typically measured using Likert-scale questions administered before and after the warning intervention, allowing researchers to identify user belief changes. While widely adopted due to its ease of measurement through self-reported user studies, this metric has limitations, as it often overlooks other important factors such as behavioral responses, emotional impact, and users' trust in the warning interventions. 
\item Perceived Objectivity: The metric assesses users' perception of a post's neutrality or impartiality. It measures the extent to which warning interventions shape users' views on the post's credibility and bias. Research by Jia et al.,~\cite{jia2022understanding} suggests that perceived objectivity can significantly shape how users process and engage with misinformation warnings. 

In this study, the researchers measured this metric by asking participants to rate their agreement with statements related to the warning interventions' objectivity, political bias, and mechanical neutrality. These responses were then analyzed using pairwise comparisons across different misinformation warning types to identify significant differences in perceived objectivity between groups.

However, a significant limitation of this metric is its reliance on self-reported data, which may be susceptible to user bias. Additionally, external factors, such as personal beliefs or ideologies unrelated to the content, could influence participants’ responses, potentially affecting the accuracy of the metric.

\item Perceived credibility: The perceived credibility metric assesses the degree to which individuals accept or believe information, regardless of whether it is labeled misinformation~\cite{luo2022credibility}. This metric gauges how warning interventions influence users' trust in the content. It serves as a major determinant of whether people will accept or reject the information they encounter~\cite{luo2022credibility}. 

This metric was measured in this study using a self-reported Likert scale, where participants rated the accuracy of news posts in their feed on a scale ranging from ``definitely fake'' to ``definitely real''

\end{itemize}

The advantages and limitations of commonly used Trust and Credibility metrics in the literature are summarized in Table \ref{tab:tab3}.

\begin{table}[h]
\caption {Trust and Credibility Metrics}
\centering
\label{tab:tab3}
\begin{tabular}{|p{2cm}|p{2.5cm}|p{2.5cm}|}
\hline
\textbf{Metrics} & \textbf{Pros }& \textbf{Cons}  \\ \hline
Perceived Credibility  & Direct insight into user trust  & Subjective \\ \hline
Perceived Objectivity  & Relevant for evaluating neutrality perceptions & Very subjective \\ \hline

Perceived Accuracy  & Easy to collect via surveys & May be influenced by belief bias \\ \hline

Trust in Warning  & reveal perceived reliability of the warning & the visual appeal may affect the measurement \\ \hline
\end{tabular}
\end{table}

Trust and credibility metrics focus on individuals' internal cognitive and emotional responses to warning interventions, measuring the extent to which users trust the credibility of the flagged content and the warning system itself. Trust and credibility metrics are valuable for understanding how users' perceptions of credibility shift in response to warning interventions. However, as shown in Table \ref{tab:tab3}, these metrics are usually based on user self-reported data through a questionnaire or interview~\cite{jia2022understanding}\cite{seo2024reliability}, making them vulnerable to social desirability biases. Moreover, trust and credibility are highly subjective, influenced by users' ideological worldviews or political inclinations~\cite{wu2019misinformation}. A significant limitation of these metrics is their inability to differentiate between trust in the content and the warning system. Users may distrust the content labeled as misinformation while simultaneously becoming skeptical of the warning intervention, particularly if they perceive the warning as biased~\cite{saltz2021encounters}.

\subsection{Usability Metrics}
Usability metrics assess how users perceive and interact with warning interventions, emphasizing ease of use, accessibility, and the overall user experience. Distinct from other metric categories, usability metrics emphasize the overall user experience and ease of interaction with the warning intervention. In the present classification, this category encompasses the following metrics:  
\begin{itemize}
\item Perceived usefulness: The perceived usefulness metric is employed in the context of misinformation warnings to evaluate how beneficial users find the intervention in detecting or avoiding misinformation. It measures the extent to which users perceive the intervention as helpful~\cite{jahanbakhsh2024browser}. Perceived usefulness is a crucial metric as it provides insights into user satisfaction, engagement, and trust in the warning intervention. As such, it is a valuable tool for assessing the overall effectiveness of misinformation warnings.
In existing literature, Pennycook et al.,~\cite{pennycook2020implied} investigated the perceived usefulness of accuracy prompts and found that users who considered the prompts useful were less inclined to share misinformation. Similarly, Jahanbakhsh et al.,~\cite{jahanbakhsh2024browser} explored the perceived usefulness of browser extensions, reporting that users found them effective in detecting misinformation. Perceived usefulness is typically assessed through self-reported Likert-scale surveys, where participants rate how helpful they find a given intervention. In this study~\cite{jahanbakhsh2024browser}, participants evaluated the usefulness of the browser extension using a Likert scale in a survey conducted by the researchers. In a related study, Clayton et al.,~\cite{clayton2020real} examined the relationship between trust and perceived usefulness of misinformation warnings, concluding that the perceived usefulness of warnings depended on the trust users had in the source.

 \item Perceived disruption: Perceived disruption metric refers to how strongly a warning intervention interrupts a user’s engagement with potentially harmful or misleading content. It captures how the warning causes users to pause, reflect, and engage critically with the content. This metric can be measured through surveys that assess users’ experiences during interactions with the warning~\cite{kaiser2021adapting}, or by tracking user behavior in experimental studies. 

 \item Accessibility: Accessibility metrics assess the ease with which individuals, regardless of physical, cognitive, or technical limitations, can interact with and comprehend a warning intervention. Considering the diversity of social media users, accessibility is a critical metric to ensure that warning interventions are inclusive and usable by all individuals, particularly those with disabilities. Evaluation of accessibility can be conducted through user testing involving individuals from varied demographic backgrounds, including those with visual impairments~\cite{sharevski2023designing}.
 
 \end{itemize}
 
The advantages and limitations of commonly used usability metrics in the literature are summarized in Table \ref{tab:tab4}.

 \begin{table}
\caption {Pros and Cons of Usability}
\label{tab:tab4}
\centering
\begin{tabular}{|p{2cm}|p{3cm}|p{3cm}|}
\hline
\textbf{Metrics} & \textbf{Pros }& \textbf{Cons}  \\ \hline
Task completion rate  & Simple and direct & It may oversimplify user behavior  \\ \hline
Cognitive load  & It highlights complexity and user strain & It is difficult to measure \\ \hline

User satisfaction  & It is simple and intuitive & It is subjective and may not reflect effectiveness \\ \hline
Aesthetic appeal  & Easy to measure through surveys & It may not reflect functional success \\ \hline
Time on task  & It helps compare different interventions & Longer time may not always mean poor usability  \\ \hline

\end{tabular}
\end{table}

 Usability metrics can be effectively measured through surveys and comprehension tests~\cite{kydd2023deep}. Given that usability is inherently dependent on individual behavior, it serves as a crucial evaluation of the effectiveness of warning interventions. However, as shown in Table \ref{tab:tab4}, these metrics may oversimplify user behavior, are difficult to measure, often subjective, and may not accurately reflect effectiveness or functional success. Additionally, longer task completion times do not always indicate poor usability. Furthermore, usability metrics often focus on users' immediate interactions, with limited attention to the long-term effects of the intervention in changing user behavior. Additionally, usability metrics such as perceived usefulness and user satisfaction are highly subjective; reports of high usefulness and satisfaction may stem from the aesthetic appeal of the warning rather than its effectiveness in altering users' beliefs about misinformation.
\subsection{Cognitive and Psychological Metrics}

Cognitive and psychological impacts play a significant role in how individuals process information. These impacts depend on how the user’s brain receives, perceives, interprets, and responds to the information. Therefore, cognitive and psychological metrics assess how users perceive, process, and respond to warnings, going beyond behavioral metrics that focus solely on observable changes in behavior. Cognitive focus on the impact of misinformation warnings on users' mental efforts to process and evaluate the credibility of content~\cite{moravec2020appealing}. In contrast, psychological impact assesses users' emotional and psychological reactions to misinformation warnings~\cite{clayton2020real}. In the present classification, this category encompasses the following metrics:

\begin{itemize}
\item Believe change: This metric is utilized in current literature to evaluate the shift in individuals' responses to warning interventions, specifically assessing the impact of such interventions in correcting misinformation. According to research by Hsu et al.,~\cite{hsu2023explanation}, this metric does not just evaluate changes in belief but also the behavioral impact of the interventions, which implies the effectiveness of the intervention in changing participants’ beliefs~\cite{hsu2023explanation}. Despite the positive reports regarding this metric in the literature, accurately measuring its effectiveness remains challenging. Additionally, while users' beliefs about the content may change, they may still share misinformation.
\item Cognitive load: Cognitive load assesses the cognitive functions of individuals when exposed to external stimuli, such as misinformation warning interventions. When encountering such warnings, these metrics evaluate various aspects of cognitive responses, including retention, critical thinking, contemplation, and skepticism~\cite{moravec2020appealing}. The primary focus is on internal processes influencing users' attitudinal changes, belief re-orientation, and behavioral modifications. These metrics provide valuable insights into how warning interventions affect individuals' perceptions of information and decision-making processes.
In their study, Moravec et al.,~\cite{moravec2020appealing} evaluated cognitive processes by testing participants' responses to different fake news warning interventions.  
 
\item Memory retention: memory retention metrics are employed within the context of misinformation warning interventions to assess whether individuals can retain both the misinformation and the warnings over time~\cite{bulevich2022witnesses}. These metrics are designed to evaluate the long-term impact of warning interventions in mitigating the recurrence of misinformation. In their research, Bulevich et al.,~\cite{bulevich2022witnesses}, observed that higher accuracy in recalling correct details indicates a reduced susceptibility to misinformation. The underlying assumption is that users who retain warnings are less likely to engage with misinformation in the future, highlighting the potential of these metrics to measure the enduring effects of such interventions. While crucial for assessing the long-term efficacy of misinformation warnings, these metrics have limitations in providing immediate insights into changes in user behavior or beliefs. For example, Bulevich et al.,~\cite{bulevich2022witnesses} found that warning interventions are less effective for highly accessible misinformation. This implies that users misinterpret information, disregarding it, diminishing the intervention's impact~\cite{bulevich2022witnesses}.
\item Attitude shift: The attitude shift metric measures changes in individuals' beliefs or opinions following exposure to a warning intervention. Conceptually, the effectiveness of a warning intervention is assessed by examining whether users become more skeptical of misinformation or modify their attitudes toward the subject matter after encountering the warning~\cite{hsu2023explanation}. This metric is typically evaluated using pre- and post-intervention surveys, where participants are asked about their beliefs regarding the misinformation both before and after exposure to the warning~\cite{hsu2023explanation}. Such measurements provide insight into the capacity of warning interventions to influence user perceptions and mitigate the effects of misinformation.

\item Cognitive resistance: The cognitive resistance metric evaluates individuals' active efforts to dismiss a warning message, particularly when it conflicts with their pre-existing beliefs. This metric is crucial for understanding how effectively warnings penetrate users' cognitive defenses and promote critical thinking~\cite{moravec2020appealing}. Cognitive resistance can be assessed through open-ended questions and cognitive response tests, which prompt users to articulate their thoughts and reactions after encountering a misinformation warning. These assessments help gauge the extent to which individuals critically engage with the warning or resist its implications~\cite{chen2022visualbubble}. 

 \end{itemize}

 The advantages and limitations of commonly used cognitive and psychological metrics in the literature are summarized in Table \ref{tab:tab5}.
 
\begin{table}
\caption {Cognitive and Psycological}
\label{tab:tab5}
\centering
\begin{tabular}{|p{2cm}|p{2.5cm}|p{2.5cm}|}
\hline
\textbf{Metrics} & \textbf{Pros }& \textbf{Cons}  \\ \hline
Belief change  & It is useful for evaluating user deep processing  & It is difficult to measure reliably  \\ \hline 

Cognitive resistance  & It helps identify user psychological pushback to warnings & Complex to measure and interpret \\ \hline

Memory recall   & It is Useful for evaluating long-term impact & May not imply belief change\\ \hline
Attitude change  & Broader than belief change & Changes may take time or be subtle \\ \hline
Attitude Shift & Useful for long-term behavior prediction & Challenging to detect and requires multiple validated measures for accurate assessment \\ \hline

\end{tabular}
\end{table}

While these metrics provide valuable insights into how users process warnings, they, like usability metrics, often emphasize short-term effects and are inherently subjective. Although they offer essential information about the effectiveness of warning interventions, these metrics are frequently challenging to measure, complex to interpret, and difficult to replicate across different contexts.

Given the challenges associated with existing metrics, particularly their subjectivity, complexity in measurement, and difficulty in replication, it is essential to establish more consistent and practical approaches for evaluating the effectiveness of misinformation warnings. To address these issues, we propose the incorporation of usability, accessibility, and contextual metrics, aiming to develop a standardized and objective framework. This approach would ensure that interventions are effective but also measurable and replicable across diverse user groups and settings. 

\section{RQ1: What are the existing metrics for assessing the effectiveness of misinformation warnings?}

We examine the existing metrics used to assess the effectiveness of misinformation warning interventions. We aim to comprehensively understand these metrics as they are applied across various studies. Table \ref{tab:tab6} summarizes the metrics identified in previous research. Notably, perceived accuracy emerges as the most frequently used metric, followed by sharing intention or likelihood of sharing. This prevalence can be attributed to these metrics' immediate insights into user responses. This section presents a summary of our findings concerning the research questions.

\begin{table*}[htbp]

\footnotesize

\centering
\caption {Existing Metrics Map with Articles }
\label{tab:tab6}
\begin{tabular}{|p{4cm}|p{11cm}|}

\hline
\textbf{Metrics/Publications} & \textbf{Definition}  \\ \hline
Share Rate~\cite{jahanbakhsh2021exploring}
  & This measures the change in the ratio of sharing true to false information. \\ \hline

  Awareness of Retraction~\cite{buczel2024forewarnings}
& This is used to measure whether participants in the study know that specific misinformation presented earlier has been retracted or corrected. \\ \hline

  Durability of the Warning~\cite{buczel2024forewarnings}

 & This is used to measure the extent to which the effects of the tested warnings persist over time. \\ \hline

 Willingness to Liking or Sharing headline~\cite{clayton2020real} & This metric measures the likelihood that participants will engage with the content by liking or sharing it. \\ \hline

 Sharing intention/Likelihood to Share~\cite{epstein2021developing}\cite{gwebu2022can}\cite{pennycook2021shifting}\cite{pennycook2020fighting}\cite{pennycook2022accuracy}\cite{pennycook2022nudging}\cite{pennycook2020implied}\cite{epstein2021developing}\cite{pillai2023explaining}\cite{lees2022twitter}\cite{seo2019trust})

  & This measures the likelihood that a participant would share the information on social media. \\ \hline

 Sharing Discernment~\cite{epstein2022explanations}
& This metric measures the difference in true and false news sharing rates. \\ \hline

 Memory Bias Awareness~\cite{grady2021nevertheless}

  & The authors use this metric to measure the study participants' ability to recognize and mitigate the influence of memory distortions caused by misinformation.
\\ \hline

Engagement (likes, shares, and reactions)~\cite{gruzd2024share}\cite{kaufman2024warning}\cite{guo2023seeing}\cite{moravec2020appealing}\cite{figl2023symbol}\cite{jia2022understanding} & This metrics are use to measures the extent to which users interact with content after an exposure to warning. Low engagement after exposure to warnings indicates the effectiveness of the warning in deterring user interaction. \\ \hline

 Perception of warnings Efficacy~\cite{guo2024all}

  & This measures how effective users believe misinformation warnings are in helping them make accurate judgments about content credibility. \\ \hline

  Cognitive Trust~\cite{gwebu2022can}

  & This metric measured the perceptions of the news' factual accuracy and overall credibility. \\
  \hline

 Trust in Information~\cite{hameleers2024alarm}

  & This metric evaluated whether warnings fostered generalized skepticism or helped participants discern trustworthy information. \\ \hline

  Perceived Risk of Misinformation~\cite{hameleers2024alarm})

  & This metric measures how much individuals believe they risk encountering or being harmed by misinformation.

 \\ \hline
 Misperception~\cite{hoes2024prominent})

  & This is used to measure the extent to which participants perceive false claims as accurate. \\ \hline

Skepticism~\cite{hoes2024prominent})

  & This is used to measure the extent to which participants doubt or distrust credible information, and increased skepticism reflects negative spillover effects of the warnings. \\ \hline

 Click through Rate (CTR)~\cite{kaiser2021adapting})

  & This metric evaluates the proportion of users who proceed beyond a warning to access potential misinformation. A high click-through rate (CTR) indicates frequent disregard for warnings, while a low CTR reflects user compliance in avoiding flagged content. \\ \hline

Alternative Visit Rate (AVR)~\cite{kaiser2021adapting})

  & This metric measures the proportion of users opting for alternative sources over flagged content. A high AVR indicates the warning's effectiveness in redirecting users to reliable information.\\
 \hline
 Behavioral Change~\cite{kaiser2021adapting}\cite{donabauer2024empowering}\cite{sharevski2022meaningful}

  & this metric measures whether the warning interventions alter users' behavior in avoiding misinformation. 
 \\ \hline

Perceived Believability~\cite{jia2022understanding}\cite{moravec2020appealing}\cite{figl2023symbol}

  & This metric assesses how credible and believable the participants find a post. A lower value of this metric reflects the effectiveness of the intervention.  \\ \hline
  
 Perceived Effectiveness~\cite{sharevski2022meaningful}\cite{moravec2020appealing}) & This metric seeks the Participants' opinion on whether the warning helps them identify and avoid misinformation.  \\ \hline

Perceived credibility~\cite{koch2023effects} & This metric measures how believable users find a piece of fake news or misinformation. A low perceived credibility indicates the effectiveness of warning interventions. \\ \hline

 Engagement Likelihood~\cite{koch2023effects}\cite{gwebu2022can} & This assesses participants’ likelihood of interacting with misinformation content. A low score implies reduced user interaction with false content.
 \\ \hline

     Perceive accuracy~\cite{jahanbakhsh2021exploring} \cite{kirchner2020countering}\cite{chen2022visualbubble} \cite{clayton2020real}\cite{grady2021nevertheless}\cite{gruzd2024share}\cite{guo2024all}\cite{hoes2024prominent}\cite{donabauer2024empowering}\cite{guo2023seeing}\cite{hsu2023explanation}\cite{jia2022understanding}\cite{pennycook2021shifting}\cite{pennycook2020fighting}\cite{pennycook2020implied}\cite{epstein2021developing}\cite{sharevski2021misinformation}\cite{lees2022twitter}\cite{seo2019trust}

  & This metric measures the extent to which participants believe a post is accurate. A Lower perceived accuracy indicates the effectiveness of the warning in reducing belief in misinformation. \\ \hline
\end{tabular}
\end{table*}

\section{RQ2: What are the challenges and opportunities of existing metrics to measure the effectiveness of misinformation warnings?}

\subsection{Challenges}

\begin{figure}[htbp]
\centerline{\includegraphics[width=8cm,height=7cm]{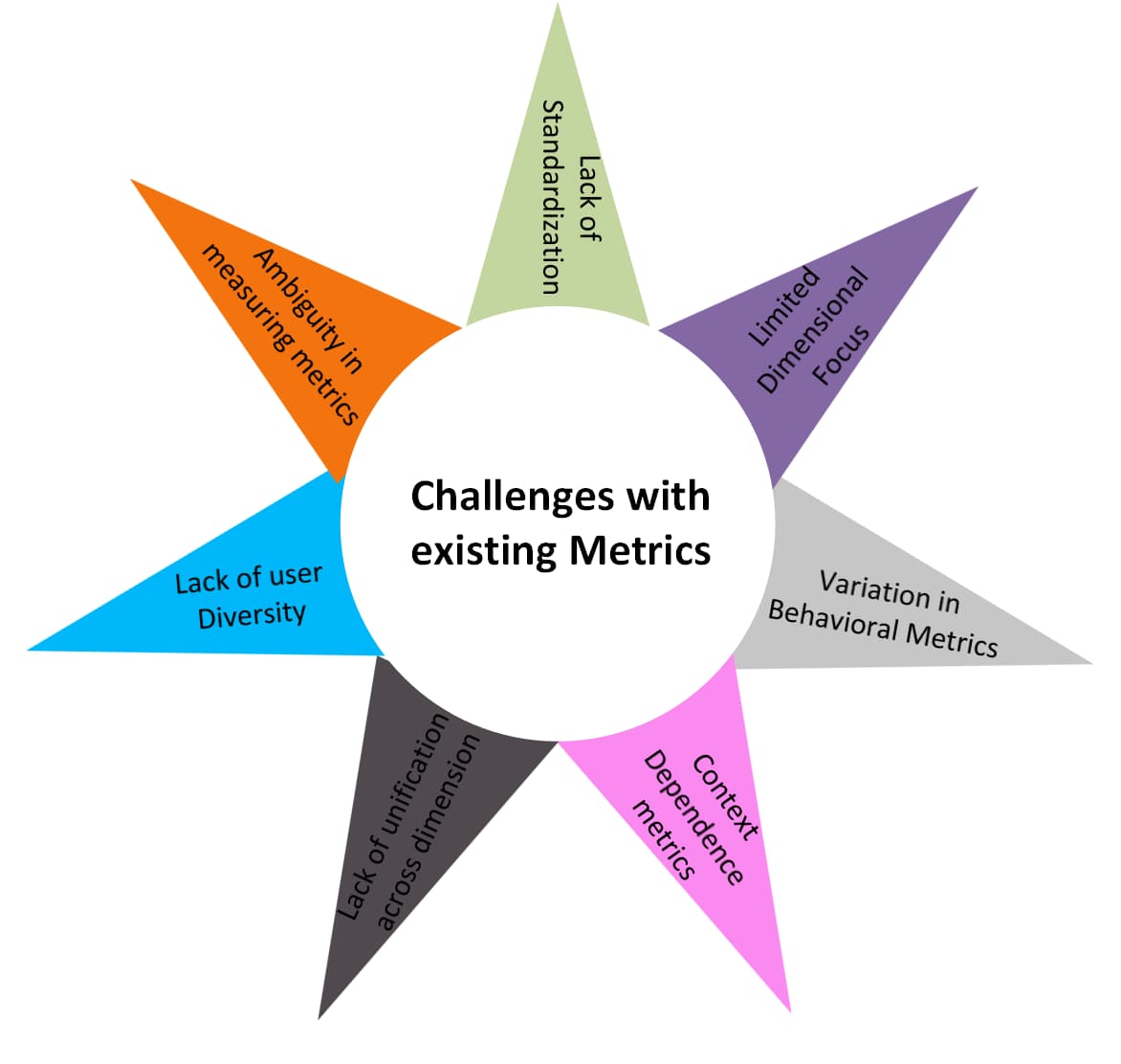}}

\caption{Challenges with Existing Metrics
}
\label{fig2}
\end{figure}

Warning interventions are increasingly recognized as a critical strategy for mitigating the spread of misinformation on social media and other digital platforms. To assess the effectiveness of these interventions, researchers have utilized various metrics designed to evaluate their impact on users' behaviors and beliefs. However, the reviewed studies revealed several challenges associated with these metrics. This section outlines and critically examines the key challenges identified in the existing metrics. Figure \ref{fig2} provides a summary of the challenges discussed.

\begin{itemize}
    \item Variation in behavioral metrics: Behavioral metrics, such as click-through rates (CTR), share rates, and engagement with alternative content among others~\cite{kaiser2021adapting}\cite{jahanbakhsh2021exploring}\cite{seo2024reliability}\cite{greenberg2024heed}\cite{jia2022understanding}, have been used in the existing literature to gauge how misinformation warnings affect users’ actions. These metrics are commonly adopted in studies assessing the immediate impact of warnings on user behavior~\cite{jahanbakhsh2021exploring}. For instance, some studies measure whether users click on or share content after encountering a warning~\cite{jahanbakhsh2021exploring}\cite{jia2022understanding}. In contrast, others focus on the extent to which users seek additional information after viewing a warning~\cite{kaiser2021adapting}\cite{kirchner2020countering}. However, the lack of consistency in how these behaviors are quantified and reported limits the comparability of findings. In addition, these metrics often fail to explain the cognitive processes behind user decisions. Furthermore, these metrics might not account for subtle behaviors, such as selective attention to warnings, or long-term impacts on users' susceptibility to misinformation.
    \item Lack of Standardization: The lack of standardized metrics presents a significant challenge in evaluating the effectiveness of warning interventions across studies. Scholars have utilized different metrics and assessment criteria, complicating the comparison of findings and the ability to draw broadly generalizable conclusions. For example, while some studies emphasize behavioral changes, such as reduced sharing of misinformation, others prioritize user engagement or cognitive aspects, including shifts in belief or the usability of warnings. This inconsistency challenges the development of a unified framework for assessing the effectiveness of warning interventions.
    \item Context Dependence: Existing metrics often fail to adequately account for the role of context in determining the effectiveness of warning interventions. Most existing studies employ generic metrics for evaluating warning interventions, neglecting the peculiar dynamics of each platform and their influence on user behavior and perception. For example, on Twitter, metrics may focus on retweets, likes, and quote tweets, while on Facebook, comments and shares could be more relevant. Similarly, platforms like Instagram, TikTok, and LinkedIn each exhibit uniqueness that influences user interactions and engagement.

    A warning intervention's effectiveness can vary significantly based on contextual factors such as the medium of communication and user engagement patterns. Metrics that perform well on one platform may fail to capture relevant outcomes on another, leading to inconsistent and incomplete assessments. Addressing these contextual variations is essential for accurately evaluating the impact of warning interventions across diverse platforms.
    \item Lack of diversity: A significant limitation of existing metrics is their lack of inclusivity, particularly in addressing the needs of visually impaired or blind users~\cite{sharevski2023designing}. Most metrics are designed with fully sighted populations in mind, disregarding how individuals with impairments respond to warnings or perceive their effectiveness. These users often rely on various assistive tools, such as screen readers or haptic feedback, which may not align with the parameters of existing evaluation metrics.

    The absence of specialized metrics for such populations creates a gap in understanding the effectiveness of warning interventions, possibly limiting valuable insights into the usability and accessibility of these interventions across diverse groups. In addition to limiting the generalization of findings, it also impedes the development of universally accessible misinformation interventions.

\item Limited dimensional focus:
Most studies rely on single-dimensional metrics, such as accuracy perception or sharing rates, failing to capture the multifaceted nature of misinformation interventions.
\item Ambiguity in measuring metrics: Some trust metrics and perceived credibility are often measured using inconsistent terminologies and methodologies, making it challenging to compare research outcomes.
\item Lack of unification across dimensions: Existing metrics fail to integrate cognitive, behavioral, usability, and trust dimensions into a cohesive evaluation framework. This fragmented approach limits the comprehensive assessment of intervention effectiveness.
\end{itemize}

\subsection{Opportunities}

As misinformation remains a public concern, the design, development, and implementation of robust metrics to evaluate misinformation warning interventions is essential. Existing evaluation metrics are inconsistent and have a limited dimensional focus, failing to capture the holistic impact of interventions on user behavior, trust, usability, and cognitive processing. Emerging prospects in this field should focus on the following:

\begin{itemize}
    \item Multidimensional Metrics:  The complexity of misinformation propagation necessitates a comprehensive approach to evaluating the effectiveness of warning interventions. Existing metrics often assess dimensions in isolation, failing to account for the interplay between behavioral, trust, usability, and cognitive factors. To address this limitation, multidimensional metrics that integrate these dimensions are essential. By assessing these aspects simultaneously, researchers and practitioners can understand an intervention's effectiveness holistically, ensuring that it addresses the multifaceted challenges associated with misinformation mitigation.
    \item Inclusive Warning Metrics: Existing warning metrics predominantly rely on visual cues, often overlooking the specific needs of visually impaired individuals~\cite{sharevski2023designing}, who are equally vulnerable to misinformation as their sighted counterparts. Developing specialized metrics tailored to visually impaired users can enhance inclusivity and contribute to more equitable and effective warning interventions for all user groups.
    \item Adaptive Evaluation: The impact of misinformation varies among individuals, influenced by factors such as educational status, prior knowledge, exposure, and cognitive biases. However, existing metrics often fail to account for these differences, potentially diminishing their effectiveness. To address this limitation, it is essential to build adaptive mechanisms that dynamically adjust metrics in real-time based on user interaction and feedback. For example, if a user constantly ignores warnings, the system should identify and assess alternative modalities, thus, optimizing the effectiveness and design of the intervention.

   \item Toward a Standardized Metrics: The lack of standardized metrics for evaluating the effectiveness of misinformation warnings remains a significant challenge in current studies. Existing research frequently utilizes customized or context-specific metrics, making it difficult to compare findings across studies and to identify the most effective measures for assessing the impact of misinformation warnings. To address this issue, future research should prioritize the development of standardized metrics that enable consistent evaluation of the effectiveness of warning interventions across various platforms and populations~\cite{altay2023headlines}.

\end{itemize}
\section{CONCLUSION AND LIMITATION}

This review highlights the current state of metrics for evaluating misinformation warnings, identifying key challenges and opportunities. Assessing the effectiveness of misinformation warning interventions requires robust and multidimensional metrics. While existing metrics provide valuable insights, they face challenges such as a lack of standardization, context dependence, and insufficient consideration of user diversity. Addressing these limitations and exploring opportunities for improvement, future research should aim to develop unified frameworks that facilitate a holistic assessment of intervention effectiveness. Such frameworks should ensure that interventions are effective, user-friendly, and reliably measurable across diverse contexts. This review is limited by its inclusion criteria, focusing exclusively on warning interventions, which may influence the findings. Future research should broaden the scope to include other interventions and prioritize the development of standardized frameworks.

\bibliographystyle{IEEEtran}
\bibliography{Ref}

\vspace{12pt}

\end{document}